# Direct visualization of the impurity occupancy roadmap in Ni-substituted van der Waals ferromagnet $Fe_3GaTe_2$

Jian Yuan,[1,†] Haonan Wang,[2,†] Xiaofei Hou,[1,†] Binshuo Zhang,[1,†] Yurui Wei,[3] Jiangteng Guo,[4] Lu Sun,[3] Zhenhai Yu,[1] Zhikai Li,[1] Xiangqi Liu,[1] Wei Xia,[1,7] Xia Wang,[1,6] Xuerong Liu,[1] Yulin Chen,[1,8] Shihao Zhang,[5,*] Xuewen Fu,[4,*] Ke Qu,[2,*] Zhenzhong Yang,[2] and Yanfeng Guo[1,7,*]

[1]School of Physical Science and Technology, ShanghaiTech University, Shanghai 201210, China

[2] Key Laboratory of Polar Materials and Devices (MOE), Ministry of Education, Shanghai Center of Brain-inspired Intelligent Materials and Devices, Department of Electronics, East China Normal University, Shanghai 200241, China

[3]School of Information Science and Technology, ShanghaiTech University, Shanghai 201210, China

[4]Ultrafast Electron Microscopy Laboratory, Key Laboratory of Weak-Light Nonlinear Photonics (Ministry of Education), School of Physics, Nankai University, Tianjin 300071, China

[5]School of Physics and Electronics, Hunan University, Changsha 410082, China

[6]Analytical Instrumentation Center, School of Physical Science and Technology, ShanghaiTech University, Shanghai 201210, China

[7]ShanghaiTech Laboratory for Topological Physics, ShanghaiTech University, Shanghai 201210, China

[8]Clarendon Laboratory, Department of Physics, University of Oxford, Oxford OX1 3PU, United Kingdom

**Impurity substitution is a general strategy to study the intrinsic properties of a quantum material. However, when the target element has more than one Wyckoff position in the lattice, it is a big challenge but with extreme necessity to**

**know the exact position and order of the occupancy of impurity atoms. Via comprehensive experimental and theoretical investigations, we establish herein the roadmap for Ni substitution in $Fe_3GaTe_2$, a van der Waals ferromagnet with the Curie temperature $T_C$ even reaching ~ 380 K. The results unambiguously reveal that in $(Fe_{1-x}Ni_x)_3GaTe_2$, Ni atoms initially form an van der Waals interlayer gap Ni3 sites when $x < 0.1$, and then gradually occupy the Fe2 sites. After replacing the Fe2 sites at $x$ of ~ 0.75, they start to substitute for the Fe1 sites and eventually realize a full occupation at $x = 1.0$. Accordingly, $T_C$ and saturation magnetic moments of $(Fe_{1-x}Ni_x)_3GaTe_2$ both show nonlinear decrease, which is tightly tied to the Ni occupancy order as well as the different roles of Ni3, Fe1 and Fe2 sites in the spin Hamiltonian. The results not only yield fruitful insights into the essential roles of different Fe sites in producing the above room temperature high $T_C$, but also set a paradigm for future impurity substitution study on other quantum materials.**

[†]These authors contributed equally to this work:
Jian Yuan, Haonan Wang, Xiaofei Hou and Binshuo Zhang.

[*]Correspondence:
S.H.Z.(zhangshh@hnu.edu.cn), X.W.F. (xwfu@nankai.edu.cn),
K.Q.(kqu@chem.ecnu.edu.cn), and Y.F.G. (guoyf@shanghaitech.edu.cn)

**INTRODUCTION**

Impurity substitution in a solid can yield fruitful insight into the mechanism underlying the physical properties. This strategy has played crucial roles in understanding the pairing symmetry of various superconductors. According to the Anderson's theorem [1], nonmagnetic impurity substitution in a superconductor with an isotropic superconducting gap influences the Cooper pairs negligibly, but can break them in a superconductor with an anisotropic gap. As a sharp contrast, magnetic impurities can cause pair-breaking effect regardless of the gap type. Guided by this theory, the $Zn^{2+}$ with a tightly closed $d$ shell was widely used as an ideal impurity to study the pairing mechanisms of both cuprate and iron-based superconductors [2-13]. Besides, the impurity substitution was also used as an approach to investigate the origin of many-body interactions in cuprate superconductors, because the substitution effect on the low-energy dynamics could be viewed as a magnetic analogue of the isotope effect [14].

The impurity substitution strategy is also widely adopted to study the recently emerged van der Waals (vdW) magnets $Fe_nGeTe_2$ ($n$ = 3, 4, 5) and $Fe_3GaTe_2$ [15-23]. The inherent magnetocrystalline anisotropy in the vdW magnets stabilizes the magnetic order against finite temperature, thus violating the Merin-Wagner theorem and realizing long-range magnetic order in a broad range of vdW magnets in mono- or few-layer form, such as in $CrI_3$, $Cr_2Ge_2Te_6$, $Cr_2Si_2Te_6$, $VSe_2$, and $MnSe_2$ [24-29], etc. The itinerant ferromagnets $Fe_nGeTe_2$ ($n$ = 3, 4, 5) are characterized by the remarkably high Curie temperature $T_C$ and therefore have been subjected to intensively investigations [30-33]. Interestingly, the $T_C$ of $Fe_nGeTe_2$ is tunable by controlling the Fe content [34-36], suggesting the pivotal role of Fe in the magnetic exchange. However, the multiple Fe Wyckoff sites in the lattice of $Fe_nGeTe_2$ give rise to complicated magnetic orders, which makes the understanding about the spin Hamiltonian extremely difficult. In $Fe_3GeTe_2$ and $Fe_5GeTe_2$, the substitution of Ni and Co shows distinct behaviors, where Co and Ni substitutions in $Fe_3GeTe_2$ gradually suppress the ferromagnetic (FM) order, while Ni substitution in $Fe_5GeTe_2$ significantly enhances the $T_C$ even up to ~ 480 K and Co substitution results in an antiferromagnetic (AFM) state [15-22]. However, with these substitutions, it remains unknown that whether the impurities enter onto the Fe sites or not, and what is the order for impurities to occupy the different Fe sites, thus hindering a direct understanding about the delicate roles of the impurities and the different Fe sites in the complex magnetic exchanges.

The recently emerged $Fe_3GaTe_2$, which is isostructural with $Fe_3GeTe_2$, has a very high $T_C$ of ~ 380 K [37]. In the centrosymmetric crystal structure, the Dzyaloshinskii-Moriya interaction (DMI) caused by the introduction of Fe deficiency and hence the spatial inversion symmetry breaking, rather than the competition between magnetic dipole interaction and strong perpendicular uniaxial anisotropy, is suggested as the driven force for the observed room-temperature Néel-type skyrmions [38, 39]. Regarding the intriguing magnetic properties related to the Fe sites, impurity substitution would yield valuable insights. A very recent impurity study work unveiled that a small amount of Co or Ni substitution in $Fe_3GaTe_2$ results in $T_C$ suppression. Specifically, the Co substitution tunes the FM state into an AFM one and then to the spin-glass state, while Ni substitution tunes the FM state into the spin-glass state [40].

In this work, by employing first-principles calculations, Cs-corrected scanning transmission electron microscopy (STEM), Lorentz transmission electron microscopy (L-TEM), magneto-optical Kerr effect (MOKE) microscopy and magnetization measurements, we established a clear roadmap for the Ni occupancy in $(Fe_{1-x}Ni_x)_3GaTe_2$ with $x$ up to 1.0.

**RESULTS AND DISCUSSION**

As is schematically drawn in **Fig. 1**a, $Fe_3GaTe_2$ crystallizes into a hexagonal structure (space group: $P6_3/mmc$) with the lattice parameters $a = b = 3.9860$ Å, $c = 16.2290$ Å, $\alpha = \beta = 90°$, and $\gamma = 120°$. In this structure, the vdW gap is between two adjacent Te atoms, and there are two Fe Wyckoff positions in each layer of the lattice, which we refer to as Fe1 and Fe2. The slabs of $Fe_3GaTe_2$ are stacked along the $c$-axis with the interlayer space of ~ 0.78 nm. Our first-principles calculations indicate that on Fe1 and Fe2 sites the magnetic moments are 1.33 $\mu_B$/Fe1 and 1.98 $\mu_B$/Fe2, respectively. In this layered structure, the intralayer exchange between the nearest Fe2 along out-of-plane direction is 26.0 meV, which is the main contribution to the FM order. As a comparison, in the crystal structure of $Ni_3GaTe_2$ as shown in **Fig. 1**b, there are addition Ni3 Wyckoff sites locating in the vdW gap. The different Fe and Ni Wyckoff sites in the two compounds make the occupancy of Ni in $Fe_3GaTe_2$ very confusing. The Bragg reflection index on the XRD spectrum of $(Fe_{1-x}Ni_x)_3GaTe_2$ ($x = 0 - 1.0$) crystals confirms the (00l) orientation, and no impurity peaks were detected within the instrument resolution limit, as shown in **Fig. S1** of the Supplementary Information (SI). The peaks of XRD for each composition can be satisfactorily indexed on basis of the hexagonal structure with the space group $P6_3/mmc$, suggesting the phase purity of our specimens. As illustrated in **Fig. 1**c, the main peaks of all compositions exhibit a continuous shift to higher angle with Ni substitution till $x = 0.7$, unveiling a monotonic decrease of the $c$-axis with increase of Ni. When $x > 0.7$, the change of $c$ shows almost saturation. The derived values of $c$-axis from the PXRD data analysis are summarized in **Table S**1 of SI. The compositions were measured by using energy dispersive X-ray spectroscopy (EDS), with the results shown in **Fig. S2** of SI. We thereafter use the measured substitution concentration for each sample. **Fig. 1**d shows the temperature dependent magnetization curves $M(T)$ for $(Fe_{1-x}Ni_x)_3GaTe_2$, which exhibits typical FM order with a $T_C$ at ~ 350 K, where $T_C$ is determined by the peak of $dM(T)/dT$. With the increase of $x$, $T_C$ is gradually suppressed, which can be clearly seen in **Fig. 1**f. When $x = 0.28$, $T_C$ decreases to ~ 50 K, while when the substitution is gradually increased up to $x = 1.0$, the FM order completely disappears, which is consistent with the fact that $Ni_3GaTe_2$ is essentially nonmagnetic [41]. The evolution of isothermal magnetization $M(H)$ at 2 K for various $x$ is consistent with the $M(T)$, clearly displaying the suppression of ferromagnetism, as seen in **Fig. 1**e. In **Fig. 1**f, the nonlinear evolution of $T_C$ and $M_S$ follows the same trend with $x$, which shows that the decrease of $T_C$ is initially very fast from 350 K to ~ 100 K before $x = 0.1$ and then is somewhat moderate till it is invisible. Considering this nonlinear evolution of $T_C$ and $M_S$ with $x$ and the fact that Ni$_3$GaTe$_2$ has three Ni sites, it is crucial to trace the positions of substituted Ni and study its effect on the magnetism. Furthermore, homogeneity and stability of the dopants are also necessary to be investigated.

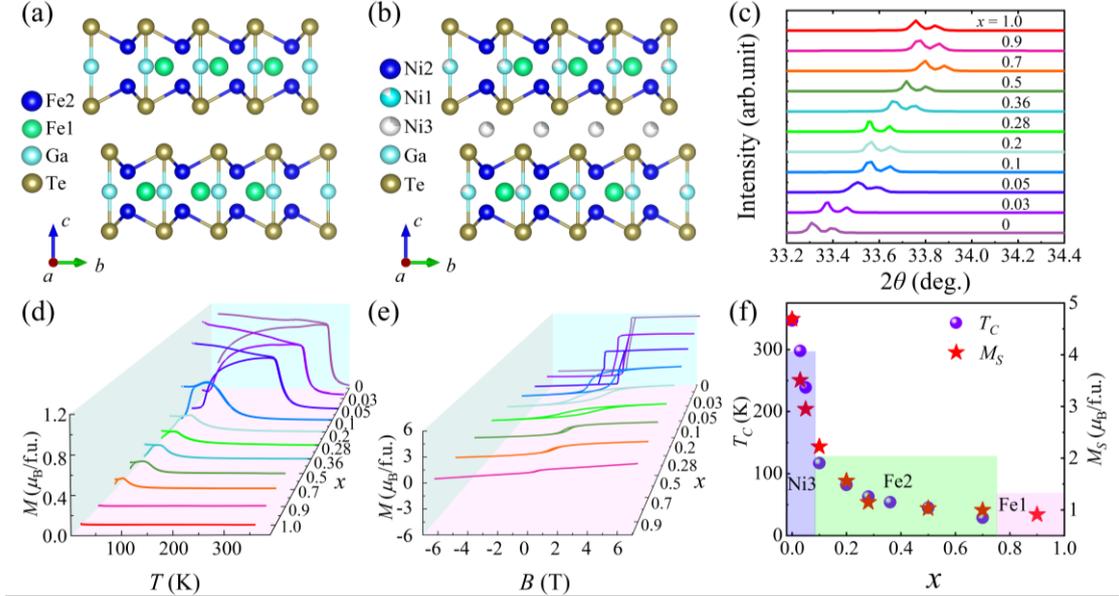

**Fig. 1. Crystal structure and magnetizations of (Fe$_{1-x}$Ni$_x$)$_3$GaTe$_2$.** (a)-(b) Schematic views of the crystal structure of Fe$_3$GaTe$_2$ and Ni$_3$GaTe$_2$ along the *a*-axis, respectively. (c) The PXRD of the (00l) orientation of (Fe$_{1-x}$Ni$_x$)$_3$GaTe$_2$. (d) Temperature dependent magnetizations of (Fe$_{1-x}$Ni$_x$)$_3$GaTe$_2$ ($x = 0 - 1.0$) under a 0.1 T magnetic field with $H \perp ab$-plane. (e) Out-of-plane isothermal magnetizations at 2 K. (f) The evolution of $T_C$ and $M_S$ at 2 K against $x$. The height and width of the rectangles represent the moment size and $x$ to achieve a full occupancy on each site, respectively.

Aberration-corrected high-angle annular-dark-field (HAADF) STEM and annular bright-field (ABF) STEM are used to obtain atomic-resolution images of Ni-substituted Fe$_3$GaTe$_2$. The image intensity in HAADF-STEM mode of an atomic column is approximately proportional to the atomic number to the power of 1.7/Z$^{1.7}$, that is, brighter spots in the images are indicative of heavier atoms in the region or similarly a greater atomic density. ABF-STEM images collect scattered electrons in a lower angle range, which can simultaneously image both light and heavy atomic species and are more sensitive to less atomic density. On the contrary to the contrast of HAADF-STEM image, dark spots are indicative of the presence of atoms in ABF-STEM mode. Therefore, by combining HAADF-STEM and ABF-STEM imaging, we can map out of the occupancy of Ni dopants within the Ni-substituted $Fe_3GaTe_2$ at the atomic level.

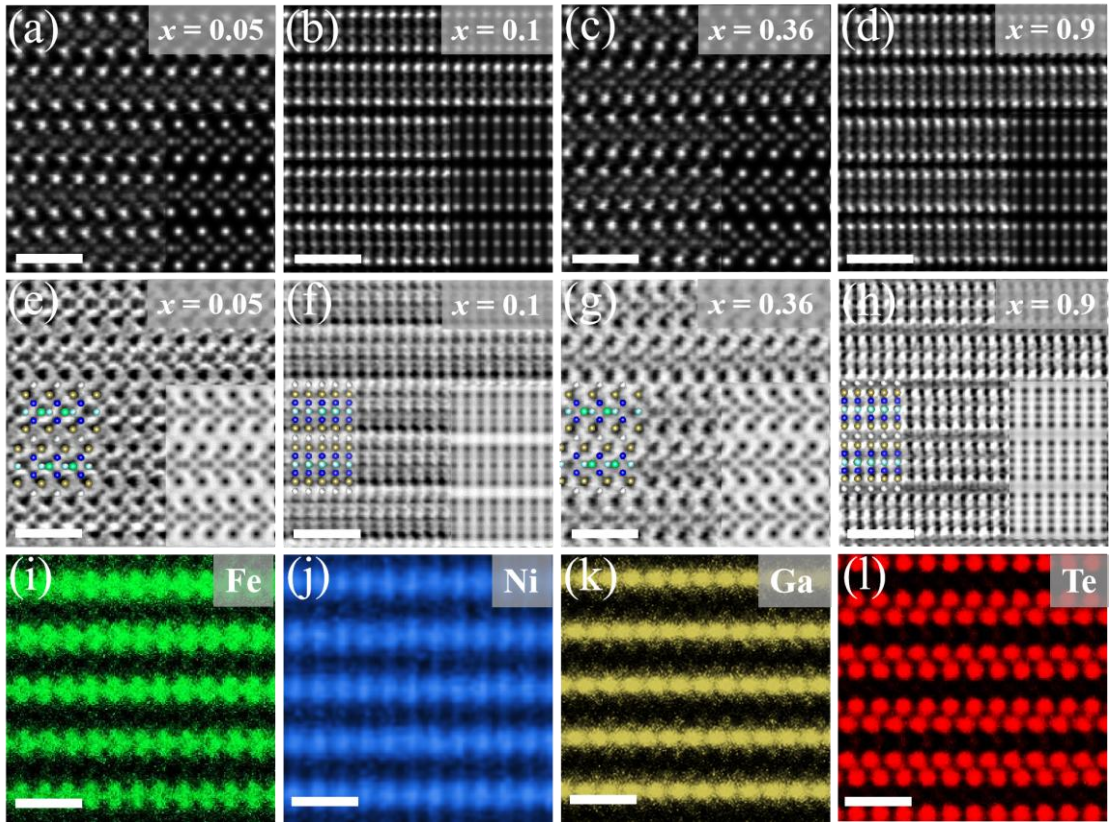

**Fig. 2. STEM images of $(Fe_{1-x}Ni_x)_3GaTe_2$ crystals with EDS mappings of Fe, Ni, Ga and Te atoms viewed along the [120] and [100] zone axes.** The HAADF-STEM images in (a)-(d) for $x =$ 0.05, 0.1, 0.36 and 0.9 and the corresponding ABF-STEM images are in (e)-(h), respectively. In (e)-(h), the cyan, blue, green, and brown balls in the crystal models represent Fe1, Fe2, Ga, and Te atoms, respectively. (i)-(l) EDS mappings of $(Fe_{0.64}Ni_{0.36})_3GaTe_2$. The scale bars in the images are 1 nm.

**Figs. 2**a-**2**d and **Figs. 2**e-**2**h show the atomic-resolution HAADF and ABF-STEM images of $(Fe_{1-x}Ni_x)_3GaTe_2$, when $x = 0.05$, 0.1, 0.36, and 0.9 respectively. Simulated HAADF/ABF-STEM images are shown as insets in the lower right corner of each figure, which are consistent with experimental images. The results demonstrate that initially, substituted Ni atoms preferentially occupy the interlayer Ni3 sites (**Figs. 2**a-**2**h). This is unsurprising, because $Ni_3GaTe_2$ possesses three distinct Ni occupation sites, unlike $Fe_3GaTe_2$ with only two Fe sites. Notably, the vdW gap between layers is partially filled with Ni atoms (0.25 occupancy), while the Ni1 site has a 75% occupancy [41]. EDS mapping results for $x = 0.36$ (**Figs. 2**i-**2**l) reveal the formation of Ni3 sites. Subsequently, additional Ni atoms preferentially occupy the intralayer Fe2 sites, as shown in the EDS mapping results for $x = 0.36$ (**Figs. 2**i-**2**l). Comparing the distribution of Fe and Ni atoms at $x = 0.36$, it is evident that Fe atoms remain in both Fe1 and Fe2 sites, while Ni atoms are restricted to Ni3 and Fe2 sites. This demonstrates that Ni atoms would actually occupy the Fe2 sites after they fill the Ni3 sites. Furthermore, with increasing Ni substitution levels, Ni atoms progressively occupy the Fe1 sites, eventually achieving complete occupancy at $x = 1.0$.

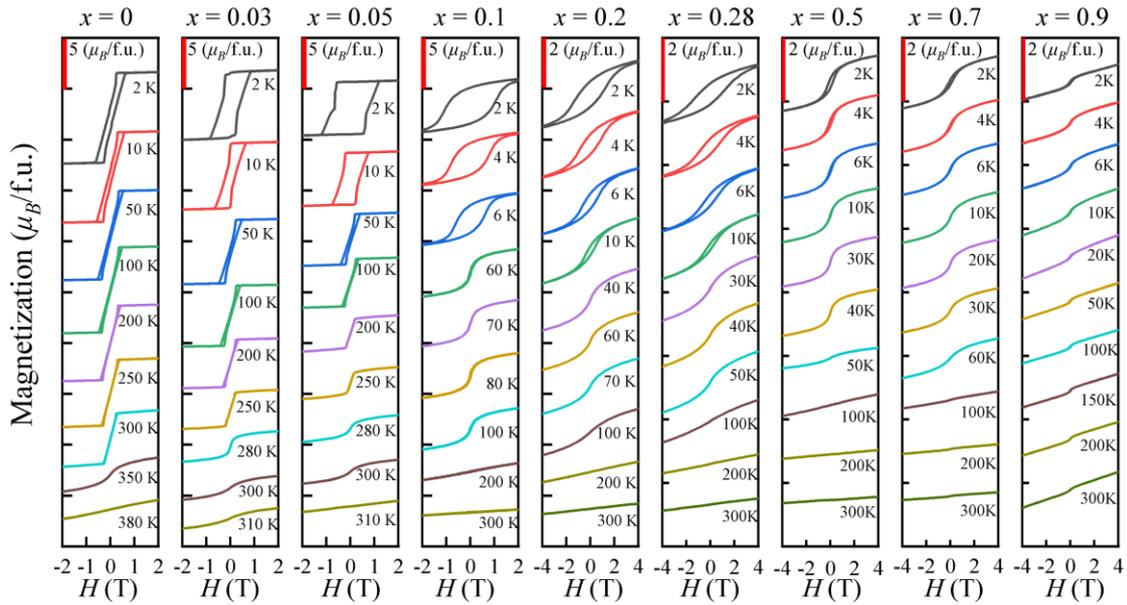

**Figure 3.** Ni substitution level *x* and temperature dependence of isothermal magnetizations measured with the magnetic field perpendicular to the *ab*-plane.

The detailed isothermal magnetizations of $(Fe_{1-x}Ni_x)_3GaTe_2$ at various temperatures measured with the magnetic field perpendicular to the *ab*-plane are presented in **Fig. 3**, which clearly show the evolution of ferromagnetism with both temperature and Ni substitution level. When $x = 0$, the hysteresis loop signifying the ferromagnetism is even visible above 300 K, which is consistent with the fact that the $T_C$ is of about 350 K. As *x* increases, the hysteresis loop becomes wider accompanied by larger saturation magnetic field and smaller saturation moment. While when $x = 0.1$, the hysteresis loop changes from square to circular shape, which then becomes narrower as *x* is further increased. The evolution of hysteresis loop is fully consistent with the STEM results, that is, when the *x* is smaller ($x < 0.1$), Ni atoms enter onto the interlayer Ni3 sites, resulting in a sudden drop of the $T_C$ and saturation moment, seen in **Fig. 1**f, while the hysteresis loop remains a rectangular shape signifying the ferromagnetism. When $x > 0.1$, Ni substitutions start to occupy the Fe2 and then the Fe1 sites and the magnetic hysteresis loops are no longer rectangular.

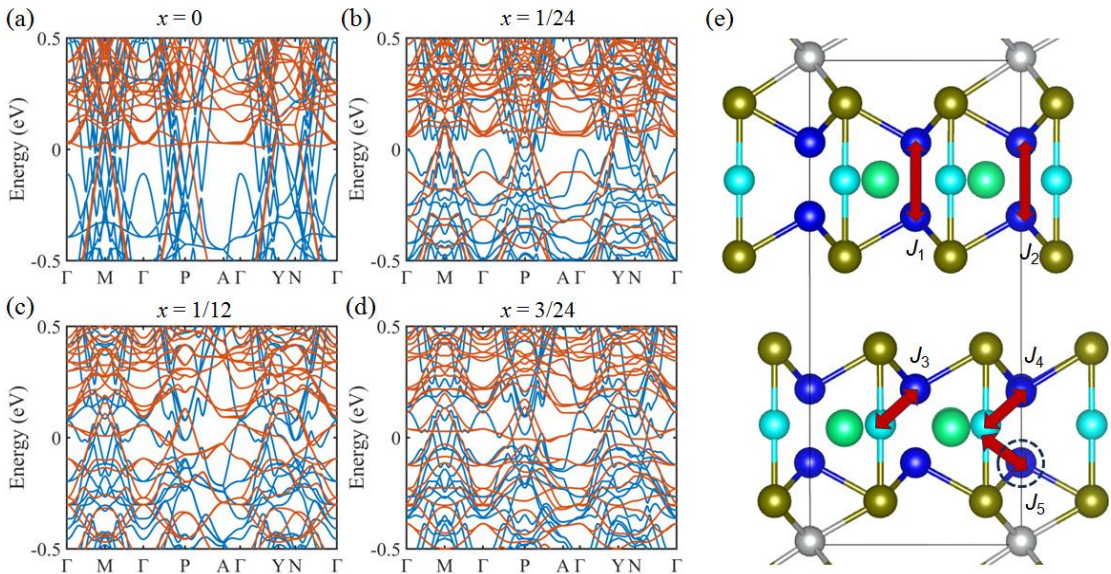

**Figure 4.** (a)-(d) The energy bands of $(Fe_{1-x}Ni_x)_3GaTe_2$ ($x = 0$, 1/24, 1/12, 3/24). The blue and red lines refer to the spin-up and spin-down energy bands, respectively. (e) The structure of $(Fe_{23/24}Ni_{1/24})_3GaTe_2$. Here the black dashed circle refers to the Fe vacancy. The different exchange parameters $J_1 \sim J_5$ are shown in the figure, where $J_1$=26.9 meV, $J_2$=17.9 meV, $J_3$=13.1 meV, $J_4$= 18.8 meV and $J_5$= 20.8 meV. The positive values of $J_1 \sim J_5$ represent the FM exchanges.

First-principles calculations can yield in-depth insights into the Ni substitution effect on the magnetism. Since the vdW interlayer gap is filled with 0.25 Ni atoms, so a supercell with 24 Fe sites (4 ×unit cell) was used in the calculations. The calculated electronic bands of $(Fe_{1-x}Ni_x)_3GaTe_2$ with $x$= 0, 1/24, 1/12, and 3/24 are presented in **Figs. 4**a-**4**d. It is apparent that the Ni substitution makes the energy bands less dispersive, and the impurity bands reduce the itinerant magnetism. During Ni substitution, the resulting Fe vacancies generate local magnetism, thereby boosting the correlation effect. Thus, $(Fe_{1-x}Ni_x)_3GaTe_2$ hold both local magnetism and itinerant magnetism due to the Fe vacancies caused by Ni doping. In the first step with the Ni substitution, i.e., when the Ni forms the interlayer Ni3 sites, the magnetic moment decreases about 4.1 $\mu_B$ per Ni atom, which indicates that substituted Ni atoms strongly weaken the itinerant magnetism. When Ni3 occupancy is larger than 0.25, the total energy of Fe2 is higher than that of Ni3 by 0.52 eV per Ni atom and higher that of Fe1 by 0.41 eV per Ni atom. Thus, the substituted Ni atoms prefer to occupy the Fe2 site in the second step. During this process, the magnetic moment decreases 2.5 $\mu_B$ per Ni atom. It indicates somewhat moderate suppression of the FM as compared to that in the first step, because the (Fe/Ni)2 site still keeps the conducting channels for itinerant electrons. This is fully consistent with our magnetization measurements. Moreover, the Fe vacancy has significant effect on the FM exchange. When $x$ = 1/24, as shown in **Fig. 4**e, assuming that in the layer without Fe vacancy, the substituted Ni atoms reduce the FM exchange interaction between the Fe1 sites near the Ni atom from 26 meV to 17.9 meV. When the layer is with Fe vacancy, the Fe vacancy remarkably influences the exchange interactions among other magnetic sites. For example, the exchange integral between Fe1 and Fe2 near the vacancy is increased to 13.1~20.8 meV. Thus, the Fe vacancies and Ni dopants simultaneously result in the inhomogeneity in $(Fe_{23/24}Ni_{1/24})_3GaTe_2$ and obviously affect the magnetic exchanges.

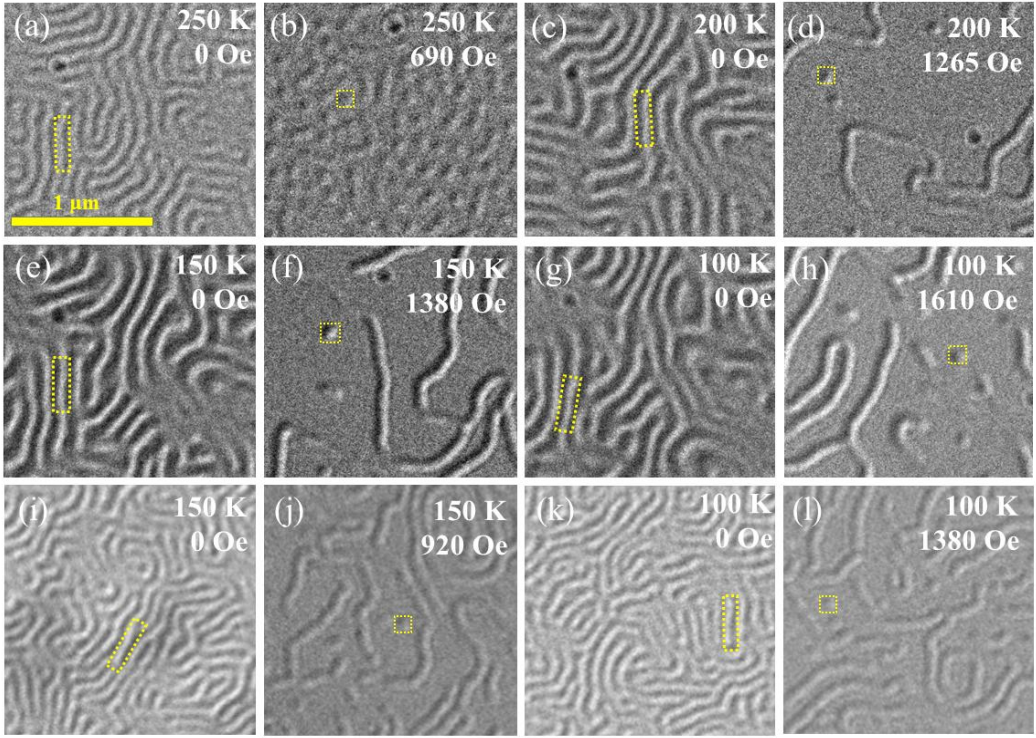

**Fig. 5. L-TEM results for $(Fe_{0.97}Ni_{0.03})_3GaTe_2$ and $(Fe_{0.95}Ni_{0.05})_3GaTe_2$ flakes.** (a)-(h) Labyrinth domain structures and magnetic skyrmions in $(Fe_{0.97}Ni_{0.03})_3GaTe_2$ with different temperatures. (i)-(l) Labyrinth domain structures and magnetic skyrmions in $(Fe_{0.95}Ni_{0.05})_3GaTe_2$ with different temperatures. The surface is within the *ab*-plane with the magnetic field perpendicular to the surface. The images are taken at $\alpha = 15°$ and $d = -3$ mm, where $\alpha$ is the angle between the sample surface and the *x* - *y* plane, *d* is the focus distance with the positive value representing the over-focus and negative value denoting the under-focus.

To see more details about the drastic suppression of FM in $(Fe_{1-x}Ni_x)_3GaTe_2$ when $x < 0.1$, we selected two representative specimens, $x = 0.03$ and $x = 0.05$, in which the saturation moments are significantly reduced, to measure their magnetic domain structures by using L-TEM. As is presented in **Figs. 5**a-**5**h, the measurements on a 100-nm-thick $(Fe_{0.97}Ni_{0.03})_3GaTe_2$ flake unveil clear labyrinthine domain structures at different temperatures below 300 K with large patterns and wide walls, indicating the presence of strong ferromagnetism. As the temperature increases, the width of domain walls gradually becomes narrower. When a magnetic field perpendicular to the *ab*-plane is applied during the measurements, the labyrinthine magnetic domains of $(Fe_{0.97}Ni_{0.03})_3GaTe_2$ gradually transform into skyrmion spin textures when the magnetic field is sufficiently large to polarize the sample, i.e. above 1600 Oe, as seen in **Figs. 5**i-**5**l. Magnetic skyrmions of $(Fe_{1-x}Ni_x)_3GaTe_2$ are observed under the magnetic fields of 1610 Oe, 1380 Oe, 1265 Oe, and 690 Oe at 100 K, 150 K, 200 K, and 250 K, respectively. At 100 K, $(Fe_{0.97}Ni_{0.03})_3GaTe_2$ and $(Fe_{0.95}Ni_{0.05})_3GaTe_2$ both have similar labyrinthine domain structures and skyrmions spin textures, but a close inspection unveils that the domain walls are wider and the skyrmions are larger when $x = 0.03$, consistent with the results of magnetization measurements. The width of domain walls can be defined as $w = 2|A/K|^{1/2}$, where *A* is exchange stiffness and *K* is uniaxial anisotropy constant [42]. The substituted Ni atoms destroy the magnetic Fe sites and decrease the total spin stiffness. Thus, the width of domain walls is decreased with Ni substitution, and the magnetic domains will become denser as well. Meanwhile, the magnetic domain structure of bulk crystal measured by using MOKE also shows the similar evolution as observed in thin flakes, as seen in **Figs. S**3 and **S**4 of SI.

## SUMMARY

To summarize, our characterizations and theoretical calculations on $Fe_{3-x}Ni_xGaTe_2$ ($x = 0 - 1.0$) crystals enable us to establish the roadmap for Ni substitution in $Fe_3GaTe_2$, which unveils three steps for Ni atoms form the Ni3 sites and occupy the different Fe sites. The substitutions are accompanied by a nonlinear change of the Curie temperature and saturation magnetic moments. In the first step, Ni atoms preferentially form the interlayer Ni3 sites, which result in a fast decrease of the Curie temperature as well as the saturation magnetic moment. Subsequently, Ni atoms enter onto the Fe2 sites and then replace the Fe1 sites, resulting in the magnetic hysteresis loop change from square to circular shape, indicating a… .. The three-step substitution mechanism is supported by the theoretical calculations. Since impurity substitution usually serves as an effective method for the study of crucial issues in many quantum materials, the results reported herein would provide the proof-of-principle elaboration of such study.

**Methods**

**Single crystal growth and characterizations**

High quality $(Fe_{1-x}Ni_x)_3GaTe_2$ ($x = 0$ - 1) single crystals were grown by using the self-flux method. High purity Fe powders (Macklin, 99.99%), Ni powders (Macklin, 99.99%), Ga lumps (Macklin, 99.999%), and Te powders (Macklin, 99.999%) in the molar ratio of 1 - $x$ : $x$ : 1 : 2 were placed into an evacuated quartz tube and sealed. The mixture was heated up to 1273 K within 10 hours, held for 20 hours and then quickly decreased down to 1153 K within 1 hour followed by slowly cooled down to 1043 K within 100 hours. Singly crystals with a typical size of $1 \times 1 \times 0.1$ mm$^3$ are obtained at the bottom of the crucible.

**Compositions and Crystallographic Phase Examinations**

Compositions of the $(Fe_{1-x}Ni_x)_3GaTe_2$ ($x = 0$ - 1) single crystals were characterized by using the energy dispersive X-ray spectroscopy (EDS, Phenom-World BV). The data obtained from measuring five points at different locations were averaged as the nominal compositions. The substitution levels $x$ of Ni refers to its real values determined by the EDS measurements. The crystallographic phase was examined by measuring the powder X-ray diffraction (PXRD) on the surface of as-grown crystals on a Bruker D8 Advance X-ray diffractometer using Cu K$\alpha$ radiation ($\lambda = 1.5408$ Å) at room temperature.

**Magnetization and electrical transport measurements**

The magnetization measurements were performed on a Quantum Design Magnetic Property Measurement System. The magnetic susceptibility $M(T)$ was measured under a 0.1 T magnetic field applied perpendicular to the *ab*-plane with the zero-field cooling (ZFC) and field cooling (FC) mode. The isothermal magnetizations at various temperatures were measured between -7 and 7 T.

**Cs-corrected scanning transmission electron microscopy Measurements**

The cross-sectional transmission electron microscopy (TEM) sample was prepared using a focus ion beam scanning electron microscopy (Helios G4 UX, Thermo Fisher). Atomic resolution high-angle annular dark-field (HAADF) and annular bright-field (ABF) scanning transmission electron microscopy (STEM) observations were performed using a 300 kV spherical aberration (Cs)-corrected STEM (JEM-ARM300F, JEOL).

**Lorentz TEM measurements and MOKE**

The magnetic domain wall contrast was observed by using a JEOL-dedicated Lorentz TEM (L-TEM, JEOL2100F). Double tilt heating holder (Gatan 652 TA) was used for high-temperature manipulation. The external perpendicular magnetic field was introduced by gradually increasing the objective lens current. The magnetic domain wall contrast at different focus was imaged under the convergent or divergent electron beam, which is introduced by the interaction of electron beam with the in-plane magnetization. To determine the in-plane magnetization distribution of a topological texture, the two sets of images with under- and over-focal lengths were recorded by a charge coupled device camera and then the high-resolution in-plane magnetization distribution map was obtained using commercial software QPt, which enabled to work out phase images and then created the magnetic field images on the basis of the transport-of-intensity equation (TIE) equation. The crystalline orientation for the grain was checked by selected-area electron diffraction. The specimen along *ab* plane and *c*-axis for L-TEM measurements were prepared via FIB milling, respectively.

By gradually changing the magnitude of the perpendicular magnetic field, the intrinsic magnetic domain structure of $x = 0.03$ and $0.05$ crystals can be visualized by measuring the MOKE (Evico Magnetics GmbH) at room temperature. During measurements, the vertical magnetic field was varied within the range of ±1 T.

**First-principles calculations**

The first-principles calculations were carried out in the framework of the local density approximation (LDA) functional [43] of the density functional theory with projector augmented wave method [44] implemented in the Vienna *ab initio* simulation package (VASP) [45]. The cutoff energy of plane wave basis set is 700 eV in all calculations. Then the Bloch states are projected to the Wannier functions [46] to build the tight-binding Hamiltonian, and we then used TB2J package [47] to calculate the exchange coupling.

**Supporting Information**

Supplementary information is available for this paper at https://xxxx.

**Data availability**

The data that support the findings of this study are available from the corresponding authors upon reasonable request.

**Competing interests**

The authors declare no competing interests.

**Acknowledgements**

The authors acknowledge the Shanghai Science and Technology Innovation Action Plan (Grant No. 21JC1402000), the National Nature Science Foundation of China (Grants No. 920651, 11934017) and National Key R&D Program of China (Grants No. 2023YFA1406100). Y.F.G. acknowledges the open research funds of State Key Laboratory of Materials for Integrated Circuits (Grant No. SKL2022) and Beijing National Laboratory for Condensed Matter Physics (2023BNLCMPKF002). S.H.Z.

was supported by the National Natural Science Foundation of China (12304217) and the Fundamental Research Funds for the Central Universities from China. The authors also thank the support from Analytical Instrumentation Center (#SPST-AIC10112914) and the Double First-Class Initiative Fund of ShanghaiTech University.

**Author contributions.**

Y.F.G. conceived the project. J.Y. synthesized the single crystals and carried out structural characterizations, measured the magnetotransport properties and analyzed the data with the help from X.F.H., B.S.Z., Z.H.Y., Z.K.L., X.Q.L., W.X., X.W. and X.R.L. J.T.G. and X.W.F. measured and analyzed the Lorentz TEM data. S.H.Z. performed the first-principles calculations. H.N.W, K.Q and Z.Z.Y. measured the atomic-resolution crystal structure by using STEM. Y.L.C. provided very useful discussions. J.Y., H.N.W., X.F.H. and B.S.Z. contributed equally to this work. Y.F.G. wrote the manuscript with input from all authors.

**References**

1. Anderson, P. W., Localized Magnetic States in Metals. *Phys. Rev.* **124**, 41 (1961).

2. Alloul, H., Bobroff, J., Gabay, M and Hirschfeld, P. J. Defects in correlated metals and superconductors. *Rev. Mod. Phys.* **81**, 452009 (2009).

3. Williams, G. V. M., Tallon, J. L. and Dupree, R. NMR study of magnetic and nonmagnetic impurities in $YBa_2Cu_4O_8$. *Phys. Rev. B* **61**, 4319 (2000).

4. Bernhard, C., Niedermayer, C., Blasius, T., Williams, G. V. M., De Renzi, R., Bucci, C. and Tallon, J. L. Muon-spin-rotation study of Zn-induced magnetic moments in cuprate high-Tc superconductors. *Phys. Rev. B* **58**, 8937 (1998).

5. Julien, M-H., Fehér, T., Horvatić, M., Berthier, C., Bakharev, O. N., Ségransan, P., Collin, G. and Marucco, J-F. $^{63}$Cu NMR Evidence for Enhanced Antiferromagnetic Correlations around Zn Impurities in $YBa_2Cu_3O_{6.7}$. *Phys. Rev. Lett.* **84**, 3422 (2000).

6. Tarascon, J. M., Wang, E., Kivelson, S., Bagley, B. G., Hull, G. W. and Ramesh, R. Magnetic versus nonmagnetic ion substitution effects on Tc in the La-Sr-Cu-O and Nd-Ce-Cu-O systems. *Phys. Rev. B* **42**, 218 (1990).

7. Armitage, N. P., Foumier, P. and Greene, R. L. Progress and perspectives on electron-doped cuprates. *Rev. Mod. Phys.* **82**, 2421 (2010).

8. Tallon, J. L. Normal-state pseudogap in $Bi_2Sr_2CaCu_2O_8$ characterized by impurity scattering. *Phys. Rev. B* **58**, 5956 (1998).

9. vom Hedt, B., Lisseck, W., Westerholt, K. and Bach, H. Superconductivity in $Bi_2Sr_2CaCu_2O_{8+\delta}$ single crystals doped with Fe, Ni, and Zn. *Phys. Rev. B* **49**, 9898 (1994).

10. Kuo, Y. K., Schneider, C. W., Skove, M. J., Nevitt, M. V., Tessema, G. X. and McGee, J. J. Effect of magnetic and nonmagnetic impurities (Ni, Zn) substitution for Cu in $Bi_2(SrCa)_{2+n}(Cu_{1-x}M_x)_{1+n}O_y$ whiskers. *Phys. Rev. B* **56**, 6201 (1997).

11. Guo, Y. F., Shi, Y. G., Yu, S., Belik, A. A., Matsushita, Y., Tanaka, M., Katsuya, Y., Kobayashi, K., Nowik, I., Felner, I., Awana, V. P. S., Yamaura, K. and Takayama-Muromachi, E. Large decrease in the critical temperature of superconducting $LaFeAsO_{0.85}$ compounds doped with 3% atomic weight of nonmagnetic Zn impurities. *Phys. Rev. B* **82**, 054506 (2010).

12. Li, J., Guo, Y. F., Zhang, S. B., Yu, S., Tsujimoto, Y., Kontani, H., Yamaura, K., and Takayama-Muromachi, E. Linear decrease of critical temperature with increasing Zn substitution in the iron-based superconductor $BaFe_{1.89-2x}Zn_{2x}Co_{0.11}As$. *Phys. Rev. B* **84**, 020513(R) (2011).

13. Li, J., Guo, Y. F., Yang, Z. R., Yamaura, K., Takayama-Muromachi, E., Wang, H. B. and Wu, P. H. Progress in nonmagnetic impurity doping studies on Fe-based superconductors. *Supercond. Sci. Technol.* **29**, 053001 (2016).

14. Terashimai, K., Matsui, H., Hashimoto, D., SATO, T., Takahashi T., Ding, H., Yamamoto T. and Kadowaki, K. Impurity effects on electron–mode coupling in high-temperature superconductors. *Nat. Phys.* **2**, 27-31 (2006).

15. Drachuck, G., Salman, Z., Masters, M. W., Taufour, V., Lamichhane, T.N., Lin, Q. S., Straszheim, W. E., Bud'ko, S. L., and Canfield, P. C. Effect of nickel substitution on magnetism in the layered van der Waals ferromagnet $Fe_3GeTe_2$. *Phys. Rev. B* **98**, 144434 (2018).

16. Tian, C. K., Wang, C., Ji, W., Wang, J. C., Xia, T. L., Wang, L., Liu, J. J., Zhang, H.-X. and Cheng, P. Domain wall pinning and hard magnetic phase in Co-doped bulk single crystalline $Fe_3GeTe_2$, *Phys. Rev. B* **99**, 184428 (2019).

17. May, A. F., Du, M.-H., Cooper, V. R. and McGuire, M. A. Tuning magnetic order in the van der Waals metal $Fe_5GeTe_2$ by cobalt substitution. *Phys. Rev. Mater.* **4**, 074008 (2020).

18. Tian, C. K., Pan, F. H., Xu, S., Ai, K., Xia, T. L. and Cheng, P. Tunable magnetic properties in van der Waals crystals $(Fe_{1-x}Co_x)_5GeTe_2$. *Appl. Phys. Lett.* **116**, 202402 (2020).

19. May, A. F., Du, M.-H., Cooper, V. R. and McGuire, M. A. Tuning magnetic order in the van der Waals metal $Fe_5GeTe_2$ by cobalt substitution. *Phys. Rev. Mater.* **4**, 074008 (2020).

20. Tian, J. C., Pan, F., Xu, S., Ai, K., Xia, T., and Cheng, P. Tunable magnetic properties in van der Waals crystals $(Fe_{1-x}Co_x)_5GeTe_2$. *Appl. Phys. Lett.* **116**, 202402 (2020).

21. Ohta, T., Kurokawa, K., Jiang, N., Yamagami, K., Okada, Y. and Niimi, Y. Enhancement of spin-flop-induced magnetic hysteresis in van der Waals magnet $(Fe_{1-x}Co_x)_5GeTe_2$. *Appl. Phys. Lett.* **122**, 152402 (2023).

22. Zhang, H. *et al.* Room-temperature skyrmion lattice in a layered magnet $(Fe_{0.5}Co_{0.5})_5GeTe_2$. *Sci. Adv.* **8**, eabm7103 (2022).

23. Zhu, K. J., Wang, M. J., Deng, Y. Z., Tian, M. L., Lei, B., and Chen, X. H. Effect of Co or Ni substitution on magnetism in the layered van der Waals ferromagnet $Fe_3GaTe_2$. *Phys. Rev. B* **109**, 104402 (2024).

24. Mermin, N. D. & Wagner, H. Absence of ferromagnetism or antiferromagnetism in one-or two-dimensional isotropic Heisenberg models. *Phys. Rev. Lett.* **17**, 1133 (1966).

25. Huang, B.*, et al.* Layer-dependent ferromagnetism in a van der Waals crystal down to the monolayer limit. *Nature* **546**, 270 (2017).

26. Gong, C.*, et al.* Discovery of intrinsic ferromagnetism in two-dimensional van der Waals crystals. *Nature* **546**, 265 (2017).

27. Lin, M.-W.*, et al.* Ultrathin nanosheets of $CrSiTe_3$: a semiconducting two-dimensional ferromagnetic material. *J. Mater. Chem. C* **4**, 315 (2016).

28. Bonilla, M.*, et al.* Strong room-temperature ferromagnetism in $VSe_2$ monolayers on van der Waals substrates. *Nat. Nanotech.* **13**, 289 (2018).

29. O'Hara, D. J.*, et al.* Room temperature intrinsic ferromagnetism in epitaxial manganese selenide films in the monolayer limit. *Nano Lett.* **18**, 3125 (2018).

30. Deng, Y. J., Yu, Y. J., Song, Y. C., Zhang, J. Z., Wang, N. Z., Sun, Z. Y., Yi, Y. F., Wu, Y. Z., Wu, S. W., Zhu, J. Y., Wang, J., Chen, X, H. and Zhang, Y. B. Gate-tunable room-temperature ferromagnetism in two-dimensional $Fe_3GeTe_2$. *Nature* **563**, 94–99 (2018).

31. Fei, Z., Huang, B., Malinowski, P., Wang, W., Song, T., Sanchez, J., Yao, W., Xiao, D., Zhu, X., May, A. F., Wu, W., Cobden, D. H., Chu, J.-H., Xu, X. Two-dimensional itinerant ferromagnetism in atomically thin $Fe_3GeTe_2$. *Nat. Mater.* **17**, 778-782 (2018).

32. Seo, J. *et al.* Nearly room temperature ferromagnetism in a magnetic metal-rich van der Waals metal. *Sci. Adv.* **17**, 8912 (2020).

33. May, A. F., Ovchinnikov, D., Zheng, Q., Hermann, R., Calder, S., Huang, B., Fei, Z., Liu, Y., Xu, X., McGuire, M. A. Ferromagnetism near room temperature in the cleavable van der Waals crystal $Fe_5GeTe_2$. *ACS Nano* **13**, 4436-4442 (2019).

34. Wu, Y. S. *et al*. Fe-Intercalation Dominated Ferromagnetism of van der Waals $Fe_3GeTe_2$. *Adv. Mater.* **35**, 2302568 (2023).

35. May, A. F., Bridges, C. A. and McGuire, M. A. Physical properties and thermal stability of $Fe_{5-x}GeTe_2$ single crystals. *Phys. Rev. Mater.* **3**, 104401 (2019).

36. Li, Z. X. *et al*. Weak antilocalization effect up to ~ 120 K in the van der Waals crystal $Fe_{5-x}GeTe_2$ with near room temperature ferromagnetism. *J. Phys. Chem. Lett.* **14**, 5456-5465 (2023).

37. Zhang, G., Guo, F., Wu, H., Wen, X., Yang, L., Jin, W., Zhang W. and Chang, H. Above-room-temperature strong intrinsic ferromagnetism in 2D van der Waals $Fe_3GaTe_2$ with large perpendicular magnetic anisotropy. *Nat. Commun.* **13**, 5067 (2022).

38. Li, Z. F. *et al*. Room-temperature sub-100 nm Néel-type skyrmions in non-stoichiometric van der Waals ferromagnet $Fe_{3-x}GaTe_2$ with ultrafast laser writability. *Nat. Commun.* **15**, 1017 (2024).

39. Hou, X. F., Wang, H. N., Zhang, B. S., Xu, C., Sun, L., Li, Z. X., Wang, X., Qu, K., Wei, Y. R. and Guo, Y. F. Room-temperature skyrmions in the van der Waals ferromagnet $Fe_3GaTe_2$. *Appl. Phys. Lett.* **124**, 142404 (2024).

40. Zhu, K. J., Wang, M. J., Deng, Y. Z., Tian, M. L., Lei, B., and Chen, X. H. Effect of Co or Ni substitution on magnetism in the layered van der Waals ferromagnet $Fe_3GaTe_2$, *Phys. Rev. B* **109**, 104402 (2024).

41. Deiseroth, H.-J., Aleksandrov, K., Reiner, C., Kienle, L., Kremer, R. K. $Fe_3GeTe_2$ and $Ni_3GeTe_2$ –Two New Layered Transition-Metal Compounds: Crystal Structures, HRTEM Investigations, and Magnetic and Electrical Properties. *Eur. J. Inorg. Chem.* **2006**, 1561–1567 (2006).

42. Yang, H.-H., Bansal, N., Rüßmann, P., Hoffmann, M., Zhang, L. C., Go, D., Li, Q. L., Haghighirad, A.-A., Sen, K., Blügel, S., Tacon, M. L., Mokrousov, Y. and Wulfhekel. W. Magnetic domain walls of the van der Waals material $Fe_3GeTe_2$. *2D Mater.* **9**, 025022 (2022).

43. Perdew, J. P. & Zunger, A. Self-interaction correction to density-functional approximations for many-electron systems. *Phys. Rev. B* **23**, 5048 (1981).

44. Blöchl, P. E. Projector augmented-wave method. *Phys. Rev. B* **50**, 17953 (1994).

45. Kresse, G. & Hafner, J. Ab initio molecular dynamics for liquid metals. *Phys. Rev. B* **47**, 558 (1993).

46. Mostofi, A. A. et al. wannier90: A tool for obtaining maximally-localised wannier functions. *Comput. Phys. Commun.* **178**, 685–699 (2008).

47. He, X., Helbig, N., Verstraete M. J., & Bousquet E. TB2J: A python package for computing magnetic interaction parameters. *Comput. Phys. Commun.* **264**, 107938 (2021).

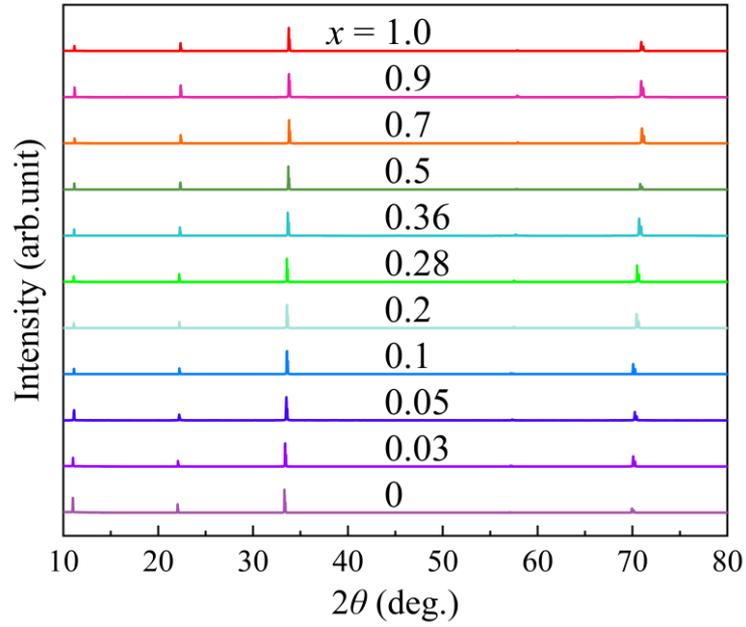

**Figure S1.** The powder X-ray diffraction patterns of the (00l) orientation of $(Fe_{1-x}Ni_x)_3GaTe_2$.

**Table S1**. The *c*-axis lattice parameters for all $(Fe_{1-x}Ni_x)_3GaTe_2$ calculated from the powder X-ray diffraction data.

| Ni doping level | *c* (nm) |
| --- | --- |
| $x = 0$ | 1.6140 |
| $x = 0.03$ | 1.6110 |
| $x = 0.10$ | 1.6055 |
| $x = 0.20$ | 1.6025 |
| $x = 0.28$ | 1.6020 |
| $x = 0.36$ | 1.5950 |
| $x = 0.50$ | 1.5930 |
| $x = 0.70$ | 1.5895 |
| $x = 0.90$ | 1.5900 |
| $x = 1.00$ | 1.5910 |

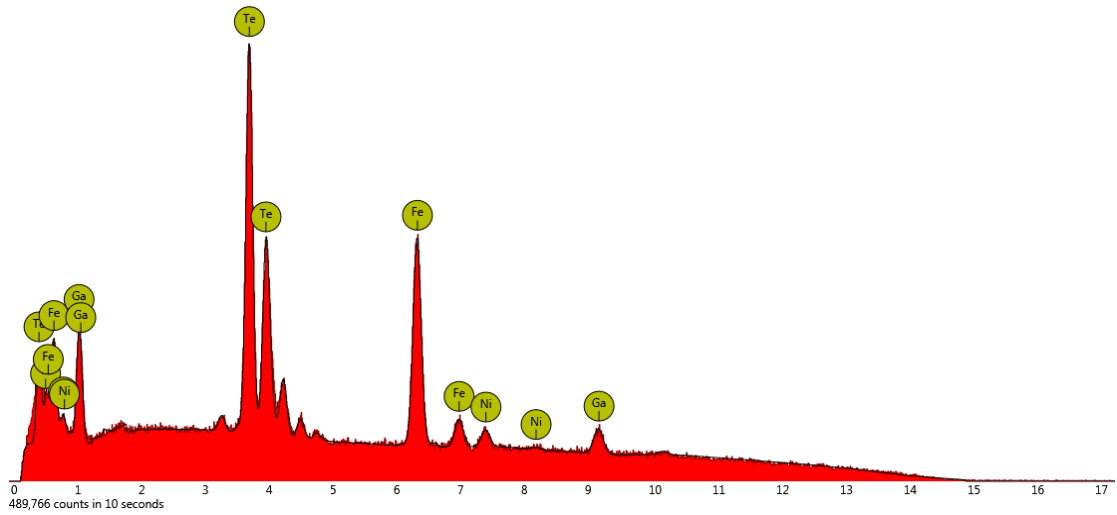

| Element Symbol | Element Name | Atomic Conc. Samp 1 | Samp 2 | Samp 3 | Samp 4 | Samp 5 | Samp 6 | Samp 7 | Samp 6 | Samp 5 | Samp 6 | Samp 7 |
|---|---|---|---|---|---|---|---|---|---|---|---|---|
| Fe | Iron | 47.36 | 49.31 | 47.48 | 41.94 | 39.59 | 33.93 | 29.72 | 16.26 | 13.09 | 3.97 | 0 |
| Te | Tellurium | 35.4 | 34.92 | 34.03 | 36.13 | 37.36 | 36.21 | 36.01 | 37.17 | 37.57 | 38.16 | 38.90 |
| Ga | Gallium | 17.24 | 14.31 | 16.05 | 16.97 | 13.7 | 16.31 | 17.91 | 18.75 | 17.79 | 19.16 | 19.68 |
| Ni | Nickel | 0 | 1.46 | 2.45 | 4.96 | 9.36 | 13.55 | 16.36 | 27.83 | 31.59 | 38.70 | 41.42 |
| The approximate value of $x$ | | $x=0$ | $x=0.03$ | $x=0.05$ | $x=0.1$ | $x=0.2$ | $x=0.28$ | $x=0.36$ | $x=0.5$ | $x=0.7$ | $x=0.9$ | $x=1$ |

**Figure S2.** The energy dispersive X-ray spectroscopy chracterizations results for $(Fe_{1-x}Ni_x)_3GaTe_2$.

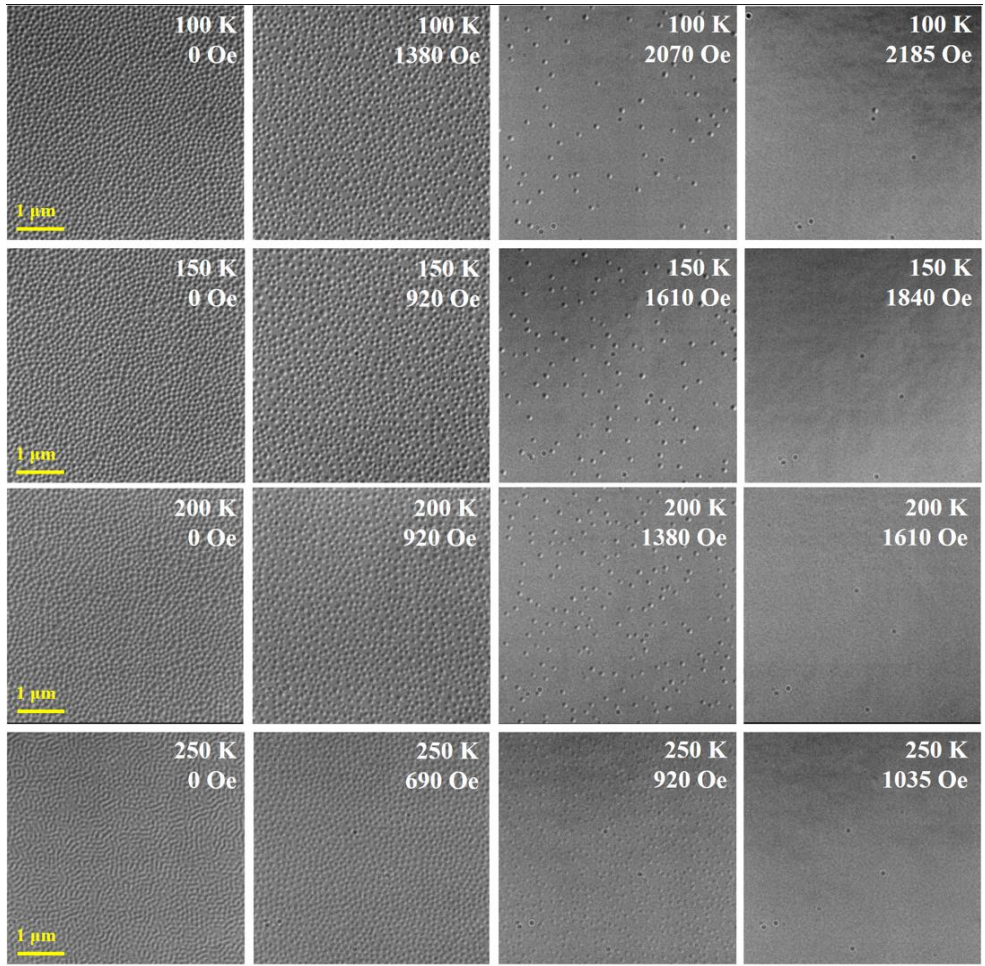

**Figure S3.** Lorentz transmission electron microscopy images of skyrmion lattice in $(Fe_{0.97}Ni_{0.03})_3GaTe_2$ after FC treatment.

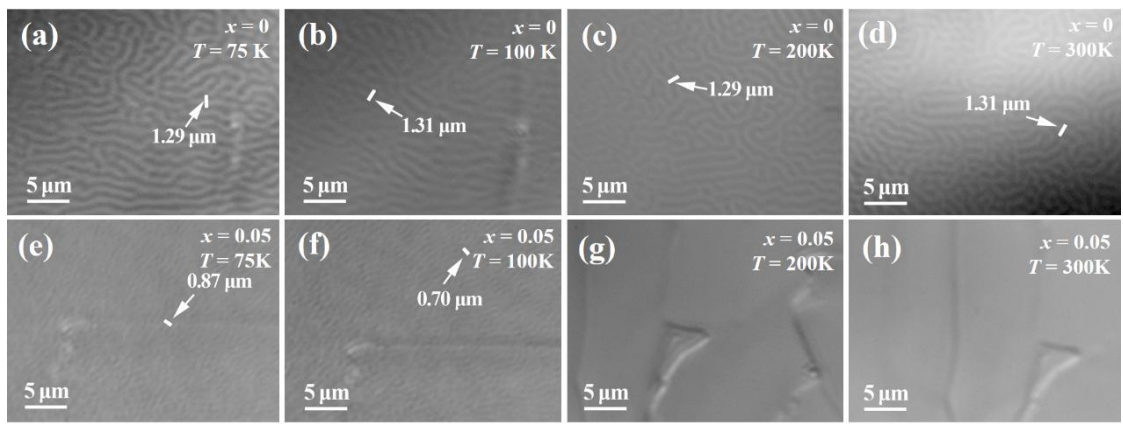

**Figure S4.** Magnetic domain imaging of $Fe_{3-x}Ni_xGaTe_2$ ($x = 0$ and $x = 0.05$) crystals by using magneto-optical Kerr effect microscopy at different temperatures without external magnetic field.